\documentclass[aps,pra,twocolumn,superscriptaddress, biblatex]{revtex4-2}

\pdfoutput=1
\usepackage[english]{babel}

\usepackage[colorlinks=true, breaklinks=true]{hyperref}
\usepackage{epsfig,graphicx}
\usepackage{indentfirst}
\usepackage{amsfonts,amssymb,latexsym,amsmath,enumerate,amsthm}
\usepackage{bm}
\usepackage{color}
\usepackage{appendix}

\usepackage{dcolumn}
\usepackage{centernot}

\usepackage{siunitx}

\usepackage{courier}

\usepackage{url}

\usepackage{physics}
\usepackage{mathtools}

\usepackage{soul}

\usepackage[sort&compress]{natbib}
\begin{document}

\title{Accurate Wichmann-Kroll vacuum polarization \\ calculations with Gaussian basis set}

\author{V.\,K.~Ivanov}
    \email[E-mail: ]{vladislav.ivanov@metalab.ifmo.ru}
\affiliation{School of Physics and Engineering, ITMO University, 
197101 St.\,Petersburg, Russia}

\author{ D.\,A.~Glazov}
\affiliation{School of Physics and Engineering, ITMO University, 
197101 St.\,Petersburg, Russia}
\affiliation{Petersburg Nuclear Physics Institute named by B.P. Konstantinov of National Research Centre ``Kurchatov Institute'',
188300 Gatchina, Leningrad region, Russia}

\author{ A.\,V.~Volotka}
\affiliation{School of Physics and Engineering, ITMO University, 
197101 St.\,Petersburg, Russia}

\date{\today}

\begin{abstract}

We apply a large Gaussian basis set of size up to $N = 500$ to calculate the all-order Wichmann-Kroll vacuum polarization correction in highly-charged ion. 
For a hydrogen-like system, this allows us to obtain an accurate correction value with a level of precision comparable to that of traditional methods.
We demonstrate that values of effects that can be calculated as the difference between two quantities, such as the nuclear model correction, converge faster than the absolute value of the correction. We discuss the basis extrapolation methods and show their use in estimating the error due to incomplete convergence. Finally, we show an application of the Gaussian basis set to calculation of the vacuum polarization screening correction in lithium-like ions.

\end{abstract}

\maketitle

\section{Introduction} 

In order to predict the spectra of highly-charged ions with high precision, one must consider the quantum electrodynamic (QED) corrections such as the electron self-energy and vacuum polarization corrections \cite{Mohr1998, Yerokhin2015, Yerokhin2019, Indelicato2019, Volotka2013, Sturm2014}.
While for light systems such as a hydrogen atom the calculations can be made using standard QED methods, for heavy highly-charged ions the non-perturbative techniques are required, 
as the constant for the interaction with the nuclear electric field is not sufficiently small \cite{Mohr1998}.
A common non-perturbative approach is to assume the static nuclear field as classic and incorporate it into the electron wave function (Furry's picture). 
In the case of vacuum polarization, the non-linear in the nucleus field (i.e. beyond the Uehling) contribution is called the Wichmann-Kroll correction, named after Wichmann and Kroll, who studied it in their work \cite{Wichmann1956}. The all-order Wichmann-Kroll correction was studied later in Refs.~\cite{Gyulassy1974, Gyulassy1975, Rinker1973, Rinker1975}. Accurate numerical results for self-energy \cite{Mohr1974} and vacuum polarization \cite{Soff1988, Manakov1989} energy shifts in hydrogen-like ions were calculated using this framework.
Since experiments nowadays have reached high levels of precision \cite{Volotka2013, Sturm2014}, higher-order corrections are of great interest. This includes the second-order radiative QED effects, and also the QED corrections to the Zeeman and hyperfine splittings \cite{Beier2000}. For example, considering such corrections in the $g$-factor allows a precise determination of the fine-structure constant \cite{Shabaev2006, Yerokhin2016}.

The finite-basis-set (or Rayleigh-Ritz) method is convenient for many atomic calculations \cite{Grant2000} such as the calculation of Zeeman and hyperfine splittings, as well as higher-order magnetic interaction effects \cite{Moskovkin2004}. Usually in this approach one uses a prescribed set of basis functions, such as B-splines \cite{Johnson1988, Sapirstein1996}. This method can also be applied to QED calculations of self-energy and photon exchange corrections, typically with dual-kinetic balance \cite{Shabaev_DKB}. 
For vacuum polarization, however, calculations are typically performed by integrating the Green's function \cite{Soff1988, Sapirstein2003}. This approach yields accurate results, and many important results have been obtained using this method (see, e.g., Ref.~\cite{Yerokhin2015}). The finite-basis-set approach was  successfully applied in its numerical form in Ref.~\cite{Persson1993}.
However, until recently, the finite-basis-set method, in which the basis functions are preset (e.g., in the form of B-splines), had not been applied successfully to vacuum polarization calculations. 
In Ref.~\cite{Salman2023},
this method was successfully applied to the vacuum polarization contribution. In this study, the authors employed a finite Gaussian basis set instead of the B-spline basis sets typically used in self-energy calculations. Later, the same basis set was applied to self-energy calculations \cite{Ferenc2025}. In Ref.~\cite{Ivanov2024} this finite-basis-set method with  a Gaussian basis set was used to evaluate the first several partial contributions of the Wichmann-Kroll correction. This allowed for a quantitative comparison of this method with older results in Ref.~\cite{Mohr1998, Persson1993}. 
The application of the Gaussian basis set method to calculate the vacuum polarization not only provides an independent computational technique, but is also promising for use in molecular QED and for evaluating complex QED diagrams with vacuum polarization loops.

These successful results have raised interest in this topic. In Ref.~\cite{Benazzouk2026} the same method was applied to calculate the Wichmann-Kroll correction, and, importantly, the complete basis extrapolation of the results was discussed.
In \cite{Mandrykina2026} a comparison was made between the finite-basis-set method and Green's function integration method (see e.g. \cite{Soff1988, Sapirstein2003}). Another recent study, Ref.~\cite{Hayat2026arxiv}, used a finite Gaussian basis to calculate Wichmann-Kroll corrections in lithium-like systems. The use of another basis set, constructed from Sturmian functions, to calculate vacuum polarization was studied in Ref.~\cite{Ivanov2025}. 

The Gaussian basis size in the above-mentioned papers was limited to $N \approx 120$. This limitation is mainly due to the fact that the Gaussian basis suffers from a linear dependence problem, which can be overcome by using increased arithmetic precision during the calculation of the vacuum polarization density. Although the convergence of the method was demonstrative, the numerical results still deviated from those of other methods. In this work, we developed a new program to calculate the vacuum polarization correction using a large basis of size $N \leq 500$. This allowed us to obtain numbers much closer to those available with standard methods such as the Green's function integration. To verify our results, we compare them with those previously obtained using the Gaussian basis set method and with those in recent papers where Green's function integration was used \cite{Sommerfeldt2026, Mandrykina2026}. We also use several nuclear models, which allows us to study the sensitivity of the results due to the nuclear charge distribution. We demonstrate that this correction converges rapidly as the basis increases. Finally, we discuss the application of the Gaussian basis set method to calculating the vacuum polarization screening effect in lithium-like ions and compare our results with those of Ref.~\cite{Persson1993, Sapirstein2003}.

Throughout the work, we use the relativistic system of units $\hbar = c = m_e = 1$, where $m_e$ is the electron mass, and the Heaviside charge unit [$\alpha = e^2/(4\pi)$, $e<0$].  

\section{Theory}

In the theoretical part of the work we closely follow our previous paper \cite{Ivanov2024}, so we present only essential details below.
In this work, we consider heavy hydrogen-like ions. Such ions are described by the Dirac equation. For a spherically symmetric system, the equation and its solution take the following form:
\begin{equation}
    h_D \phi_{n,\kappa,m_j} (\vb{x}) = E_{n,\kappa} \phi_{n,\kappa,m_j}(\vb{x})
\label{dirac_eq_2}
\end{equation}
with the Hamiltonian
\begin{align}
    h_D =& -i\boldsymbol{\alpha}\nabla + \beta + V_\mathrm{nucl}(\vb{x}), \label{wave_1}\\    
    \beta =& \gamma^0, \\
    \boldsymbol{\alpha} =& \gamma^0 \boldsymbol{\gamma} = \begin{pmatrix}
        0 & \boldsymbol{\sigma} \\
        \boldsymbol{\sigma} & 0
    \end{pmatrix},
\label{wave_2}
\end{align}
where $\gamma$ and $\boldsymbol{\sigma}$ are the Dirac and Pauli matrices, respectively. The solution of this equation is
\begin{equation}
    \phi_{n,\kappa,m_j} (\vb{x}) = \frac{1}{r}
    \begin{bmatrix}
        P_{n,\kappa}(r) \Omega _{\kappa,m_j}(\theta,\varphi) \\
        i Q_{n,\kappa}(r) \Omega _{-\kappa,m_j}(\theta,\varphi)
    \end{bmatrix},
\label{anzatz}
\end{equation}
where $P$ and $Q$ are the large and small components, respectively, of the wave function, and $\Omega _{\kappa,m_j}$ is the spherical spinor. In the case of spherical symmetry, the spherical spinors factor out, allowing the equation to be solved with only the radial coordinate.

The expression for the Wichmann-Kroll vacuum polarization charge density is well known (see e.g. Ref.~\cite{Mohr1998}).
For a spherically symmetric problem, it is convenient to solve for individual partial-wave components of the vacuum polarization charge density, i.e. for $\rho_\kappa$. 
We use the following expression for these components \cite{Salman2023}:
\begin{align}
    \rho_\kappa(\vb{x}) =& \frac{\abs{\kappa}}{2\pi}\frac{e}{2}\frac{1}  {r^2}\sum\limits_n \text{sgn}(E_{n, \kappa})\rho_{n, \kappa}(r),
    \label{rho1}
    \\ \rho_{\kappa,n}(r) &= P_{n,\kappa}^2 + Q_{n,\kappa}^2.
\end{align}
For the exact solution, the sum above should be considered symbolically, since both discrete and continuous spectra must be taken into account. In the finite-basis-set method, this sum becomes a finite sum over the spectrum obtained by a variational procedure.

We are interested in calculating the Wichmann-Kroll (WK) contribution. This part is obtained by subtracting the linear in $Z\alpha$ (Uehling) contribution from the total vacuum polarization density. The Uehling contribution is divergent; thus, the renormalized expression for the Uehling potential is usually considered separately. The expression for renormalized Uehling potential is well known and need no sophisticated techniques to evaluate. The unrenormalized charge density can be calculated using an expansion in the nuclear charge. To obtain the correct numerical result for the Wichmann-Kroll term, it is important to ensure that the $\mathcal{C}$-symmetry is preserved. Therefore, we calculate the following density \cite{Salman2023}:
\begin{equation}
\begin{aligned}
    \rho_{\kappa,\mathcal{C}}^{n \geq 3}(r,Z) &\approx \frac{1}{2}\left(\rho_\kappa(r,Z) - \rho_\kappa(r,-Z)\right) \\
    &- \frac{1}{2}\frac{Z}{\delta}\left(\rho_\kappa(r,\delta) - \rho_\kappa(r,-\delta)\right).
    \label{c_sym}
\end{aligned}    
\end{equation}

\subsection{Gaussian basis set}

We employ a finite basis set method to calculate the vacuum polarization charge density. The components of the wave function are approximated using a linear combination of the basis functions:
\begin{align}
    P_{n,\kappa}(r) &= \sum\limits_{i=1}^N v_i \pi^L_i(r) \label{basis_fncs1} ,\\
    Q_{n,\kappa}(r) &= \sum\limits_{i=N+1}^{2N} v_i \pi^S_{i-N}(r) .
    \label{basis_fncs2}
\end{align}
We use the Gaussian basis set in our calculations. The Gaussian basis set for a relativistic spherically symmetric system with a finite-size nucleus is defined as follows:
\begin{equation}
    \pi^{\tau}_i(r) = \mathcal{N} r^{l_\tau} e^{-\zeta_i r^2},
    \label{gauss_basis}
\end{equation}
\begin{equation}
    l_\tau = \abs{\kappa \pm \frac{1}{2}} + \frac{1}{2},
\end{equation}
where we denote $\tau = L, S$ symbolically for the large and small components of the wave function, respectively, $\mathcal{N}$ is a normalization constant.

To define a basis, we choose the parameters $\zeta_i$ such that they form a geometric progression:
\begin{equation}
    \zeta_i = \zeta_1 \left(\frac{\zeta_N}{\zeta_1}\right)^{\frac{i-1}{N-1}} \equiv \zeta_1 \beta^{i-1} .
    \label{zetas}
\end{equation}
The complete basis set is obtained with the following limit \cite{Klahn1985}:
\begin{equation}
    \zeta_1 \to 0^+ , \,\, \beta \to 1^+, \,\, N \to \infty.
    \label{complete_basis}
\end{equation}
In practice, approaching this limit is complicated by the numerical linear dependence of the Gaussian basis functions. Attempting to calculate the vacuum polarization correction with a numerical precision that is too low for the chosen basis parameters results in a collapse of the solution. We discussed this problem in the context of vacuum polarization calculations in Ref.~\cite{Ivanov2024}. In particular, the precision above the standard double format is necessary to obtain precise results in the case of the vacuum polarization calculations. 
The Eq.~\eqref{complete_basis} can be used to justify the basis extrapolation, as shown in Ref.~\cite{Benazzouk2026}. We discuss this approach in Sec.~\ref{sec_basis_extr}.

\section{Results}

We developed a program in the \texttt{C++} language for calculation of the Wichmann-Kroll vacuum polarization correction using the \texttt{Eigen3} and \texttt{Boost::math} and \texttt{Boost::multiprecision} libraries.
To solve the Dirac equation, we use the diagonalization method based on Cholesky decomposition. For the basis set, we choose
$\zeta_1 = \alpha^2 \cross 10^3$, $\zeta_N = \alpha^2 \cross 10^{11}$
, where $\alpha$ is the fine-structure constant. We set the value of $\alpha = 1/137.035\,999\,177$ from CODATA 2022 \cite{mohr2024codata}. For large basis set calculations, multiple precision was used. In the current version of the program, for $N \leq 300$ it runs on an ordinary 16 GB RAM laptop. For larger $N$ we use the computation server with a large RAM. The computation of the results for a given value of $\abs{\kappa}$ takes several minutes for $N \leq 200$ and several hours for larger $N$. To generate screening potentials, we used a FORTRAN program with Dual-kinetic-balanced B-splines \cite{Shabaev_DKB} code. 

We consider three nuclear models with shell, sphere and two-parameter Fermi (2pF) charge distribution. The potentials for the first two distributions are described by
\begin{equation}
    V_\mathrm{nucl}(r) = \begin{cases}
        - Z\alpha/r_n, \, &r\leq r_n \\
        - Z\alpha/r, \,\,\, &r > r_n,
    \end{cases}
\end{equation}
where $r_n = \expval{r^2}^{1/2}$ is the root-mean-square size of the nucleus, and
%
\begin{equation}
    V_\mathrm{nucl}(r) = \begin{cases}
        - Z\alpha/(2R_n) \left(3 - r^2 / R_n^2\right) , &r\leq R_n \\
        - Z\alpha/r, \,\, &r > R_n,
    \end{cases}
\end{equation}
%
where $R_n = \sqrt{5/3}\, r_n$. The Fermi charge distribution is given by
\begin{equation}
    \rho_\mathrm{nucl} (r) = \frac{\rho_0}{1 + \exp \frac{r-c}{a}} ,
\end{equation}
where $a = \frac{t}{4 \ln 3} \approx 0.5234$ fm, $t = 2.30$ fm, and $c$ 
is defined such that $\rho(c) = \rho_0 /2$ and can be found from $\expval{r^2} \approx \frac{3}{5} c^2 + \frac{7}{5} \pi^2 a^2 $, $\rho_0$ is the normalization constant. The corresponding expression for the potential can be found in \cite{Parpia1992}, we do not repeat it here since it is rather cumbersome.

To find the energy correction, we first calculate the induced charge density using multiprecision. Then we convert the result to double precision and calculate the potential $V_\text{ind}(r)$. Finally, we use the $E_\text{WK} = \bra{1s_{1/2}} V_\text{ind} \ket{1s_{1/2}}$ formula to find the energy correction. To keep the results stable up to five digits, we calculate the induced charge density up to $r = 15 \lambdabar$, where $\lambdabar$ denotes the Compton wavelength of the electron. Since calculating the electron wave function is not a challenging problem itself,
we show only the results for the electron's ground state.

\subsection{Calculations using the large basis set}

\begin{figure}[!htb]
    \centering
    \includegraphics[width=1\linewidth]{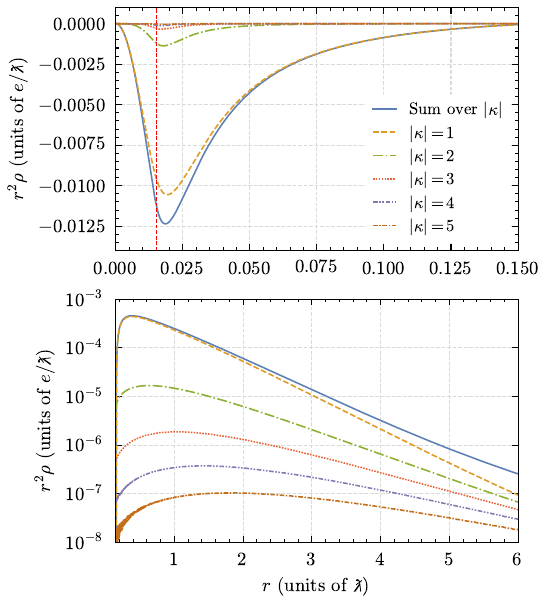}
    \caption{Radial vacuum polarization charge density partial components, calculated using the Gaussian basis set of size $N = 500$. A vertical line denotes the nuclear radius. 
    }
    \label{fig:vp_plot}
\end{figure}

In Fig. \ref{fig:vp_plot} we plot the radial vacuum polarization density $ r^2\rho_{\abs{\kappa},\mathcal{C}}^{n \geq 3}(r)$, calculated with a basis set of size $N = 500$, for $Z=92$ and the shell nucleus with $r_n = 5.871$ fm. Similarly to Refs.~\cite{Soff1988, Mohr1998}, we plot the main five partial components of the vacuum polarization density and their sum. Compared to our previous results in \cite{Ivanov2024}, this time the density curves are almost perfectly smooth at both close and far distances from the nucleus center, repeating the results of Ref.~\cite{Mohr1998}. 
If we were to plot additional partial components with higher values of $\abs{\kappa}$, they would turn out to be slightly oscillatory.
As we noted in Ref.~\cite{Ivanov2024}, this is mainly due to the small absolute value of the charge density at a given radius. For the components presented here, the oscillations would begin at a greater distance from the nucleus $r \gtrsim 10 \lambdabar$, which is also due to the small absolute value of the density.

Next, we examine how the results for the main $\abs{\kappa} = 1$ partial contribution converge as the size of the basis increases. Table \ref{tab:e_wk1} shows the values of $E_{\text{WK}}$ and the corresponding basis size $N$ and $\beta$. Since a larger basis set requires the use of higher precision, we increased it accordingly. The table lists the number of digits as defined in the \texttt{Boost::multiprecision} library. Specifically, we used the \texttt{cpp\_dec\_float\textless n\_digits\textgreater} backend, where the parameter \texttt{n\_digits} specifies the number of significant decimal digits. Where possible, the corresponding standard types are shown in parentheses. As we can see, the quadruple precision is sufficient to obtain two decimal digits of the energy shift for the main partial contribution. With $N = 500$, we reach four correct decimal digits, as confirmed by the extrapolation. For higher values of $\abs{\kappa}$, the convergence becomes slower. We discuss this below in the text.

\begin{table}[!htb]
\footnotesize
\caption{\label{tab:e_wk1}%
$E_\text{WK}$ for $\abs{\kappa} = 1$ as a function of the basis size $N$. The arithmetic precision in terms of numbers of digits is shown. Shell nucleus with $r_n = 5.751$ fm is considered.
}
\begin{ruledtabular}
\begin{tabular}{c|c|cl}
$N$  &$\beta$& $E_{\text{WK}}$, $\abs{\kappa} = 1$ & num. of digits\\

\hline
50   &1.456& 4.4886& 16 (double) \\
100  &1.205& 4.4755& 35 (quadruple) \\
150  &1.132& 4.4740& 35 (quadruple) \\
200  &1.097& 4.4736& 50 \\
250  &1.077& 4.4735& 60 \\
300  &1.064& 4.4734& 80 \\
350  &1.054& 4.4734& 90 \\
400  &1.047& 4.4734& 110 \\
450  &1.042& 4.4734& 130 \\
500  &1.038& 4.4733& 140 \\
\hline
2000\footnote{Inverse power extrapolation in $N$, Eq.~\eqref{inverse_pow}, with $i_\text{max} = 4$} & 1.009 & 4.4733 &  
\end{tabular}
\end{ruledtabular}
\end{table}

Table \ref{tab:gauss_extr} compares our present results obtained using a Gaussian basis set with $N = 300$, our previous results \cite{Ivanov2024}, which were obtained using Green's function integration and a Gaussian finite basis set method, and the results from Ref.~\cite{Benazzouk2026} obtained by the finite-basis-set calculation and basis extrapolation technique. As we can see, our recent results are close to the results obtained by the integration of Green's function in Ref.~\cite{Ivanov2024}. 
The results of the basis extrapolation of \cite{Benazzouk2026} are close to our results obtained with $N = 300$. However, such an extrapolation on the one hand has a certain degree of subjectivity, and on the other hand, it turns out to be problematic for higher values of $\abs{\kappa}$. Nevertheless, this technique can be useful to quantify the convergence of the method. We discuss this matter in detail in Sec. \ref{sec_basis_extr} below.

\begin{table}[!htb]
\footnotesize
\caption{\label{tab:gauss_extr}%
A comparison of the partial terms of $E_\text{WK}$, calculated using the finite-basis-set and Green's function methods for the ground state of electron in a hydrogen-like uranium atom with $r_n = 5.751$ fm. Our present results for Gaussian basis set with basis size of $N = 300$, as well as the results from Refs.~\cite{Ivanov2024} and \cite{Benazzouk2026} are shown.
}
\begin{ruledtabular}
\begin{tabular}{c|cccccc}
$\abs{\kappa}$ & \cite{Ivanov2024}\tablenotemark[1] & \cite{Ivanov2024}\tablenotemark[2] & { \cite{Benazzouk2026}\tablenotemark[2]}  & \cite{Benazzouk2026}\tablenotemark[3] & \cite{Benazzouk2026}\tablenotemark[4]  
& $N = 300$ 
\\

\hline
1 & 4.473 & 4.479 & 4.4746  & 4.4734& 4.4733 
&4.4734 
\\
2 & 0.394 & 0.396 & 0.3964  & 0.3939& 0.3931 
&0.3938
\\
3 & 0.081 & 0.085 & 0.0852  & 0.0806& 0.0810 
&0.0817
\\
4 & 0.024 & 0.029 & 0.0294  & 0.0246& 0.0242 
&0.0252
\\
5 & 0.009 & 0.014 & 0.0147 
& 0.0092& 0.0094&0.0099\end{tabular}
\end{ruledtabular}

\tablenotetext[1]{Green's function method}
\tablenotetext[2]{Finite Gaussian set with $N=120$}
\tablenotetext[3]{AAA extrapolation}
\tablenotetext[4]{Cubic extrapolation}

\end{table}

Although the vacuum polarization effect depends mainly on the r.m.s. radius of the nucleus, our level of precision allows us to study corrections due to the nuclear models. In Table \ref{tab:models} we present the results for the Wichmann-Kroll correction, obtained for various nuclear models and the first few partial components. As we can see, for $\abs{\kappa} = 4$ the difference between the models is already zero. Therefore, we only need the first three partial components to fully describe the model effect. We show the sum of these partial components in the second to last line of the table. We verify our results by comparing them with recent calculations in \cite{Sommerfeldt2026}. An important finding is that, although the absolute value of the correction converges slowly, the difference between the models converges much faster. In particular, this enables us to find the energy correction with a large basis set for one nuclear model and then use these numbers to find the correction for another model without performing costly calculations. We demonstrate this in Table \ref{tab:e_wk_2pf}, which shows the correction for the 2pF nuclear model. Since our results agree well with Ref.~\cite{Mandrykina2026} up to $\abs{\kappa} = 10$, we also calculate the $\kappa$-extrapolation terms to estimate the total Wichmann-Kroll correction. We ignore the $\abs{\kappa} = 11$ term since it is absent in Ref.~\cite{Mandrykina2026}. 
Using the function $\hat{E}(\abs{\kappa};\vb{a}) = a_0 \abs{\kappa}^{a_1}$, we fit the terms for $\abs{\kappa}=6-10$ and sum the extrapolated terms to $\abs{\kappa} = 40$, which yields an additional $0.00172$ eV.

Suppose that we had no reference values from other methods. Then only from the results from the Gaussian-basis-set method in Table~\ref{tab:e_wk_2pf} we could be confident in values for $\abs{\kappa} \leq 6$ (see Sec.~\ref{sec_basis_extr} for the discussion). Using these values and extrapolation for higher $\abs{\kappa}$ we find $E_\text{WK} = 4.9881$ eV.

\begin{table}[!htb]
\footnotesize
\caption{\label{tab:models}%
Results for $E_\text{WK}$ partial components for various nuclear models, $Z = 92$, $r_n = 5.8571$ fm. The differences between the shell model and sphere model, $\delta_\text{sph.}$, and the shell and Fermi models, $\delta_\text{2pF}$, are shown.
}
\begin{ruledtabular}
\begin{tabular}{c|c|S[table-format=1.6, table-number-alignment=center]S[table-format=1.6, table-number-alignment=center]S[table-format=2.5, table-number-alignment=center]S[table-format=1.6, table-number-alignment=center]S[table-format=2.5, table-number-alignment=center]}
$\abs{\kappa}$&$N$ & {shell} & {sphere} & {$\delta_\text{sph.}$} & {2pF} & {$\delta_\text{2pF}$} \\

\hline
 1&50 & 4.479964& 4.482596& 0.00263& 4.483674&0.00371\\
 &100 & 4.467509& 4.470134& 0.00263& 4.471209&0.00370\\
 &150 & 4.466038& 4.468661& 0.00262& 4.469735&0.00370\\
 &500 & 4.465326& 4.467948& 0.00262& 4.469022&0.00370\\
 \hline
2&50  & 0.417575& 0.417501& -0.00007& 0.417490&-0.00008\\
 &100 & 0.397527& 0.397441& -0.00009& 0.397427&-0.00010\\
 &150 & 0.394888& 0.394800& -0.00009& 0.394791&-0.00010\\
  \hline
3&50  & 0.110652& 0.110650& -0.00000& 0.110652&0.00000\\
 &100 & 0.086875& 0.086865& -0.00001& 0.086864&-0.00001\\
 &150 & 0.083344& 0.083331& -0.00001& 0.083329&-0.00001\\
  \hline
4&50  & 0.056019& 0.056019& 0.00000& 0.056022&0.00000\\
 &100 & 0.031391& 0.031392& 0.00000& 0.031393&0.00000\\
 &150 & 0.027208& 0.027207& 0.00000& 0.027207&0.00000\\
    \hline
total & 150 & 4.97148& 4.97400& 0.00252& 4.97506&0.00358\\
\cite{Sommerfeldt2026} & {} & {}& {}& 0.0024& {} &0.0034\\
\end{tabular}
\end{ruledtabular}
\end{table}

\begin{table}[!htb]
\footnotesize
\caption{\label{tab:e_wk_2pf}%
Partial expansion of $E_\text{WK}$, computed for shell model with basis sizes of $N=300$ and $N=500$ and for Fermi nucleus distribution with $r_n = 5.8571$ fm. Values from \cite{Mandrykina2026} are shown for comparison.
}
\begin{ruledtabular}
\begin{tabular}{c|cc
    S[table-format=1.4, table-number-alignment=center, parse-numbers=false]
    S[table-format=1.5, table-number-alignment=center]
}
$\abs{\kappa}$ &  $N=300$& $N=500$& \text{2pF}\tablenotemark[1] & {\cite{Mandrykina2026}} \\

\hline
1 &  4.4654&4.4653& 4.4690& 4.4689
\\
2 &  0.3937&0.3935& 0.3935& 0.3934
\\
3 &  0.0817&0.0815& 0.0815& 0.0814
\\
4 &  0.0252&0.0249& 0.0249& 0.0248
\\
5 &  0.0099&0.0096 & 0.0096 & 0.0096
\\
6 &  0.0047&0.0044 & 0.0044 & 0.0044
\\
7 &  0.0027&0.0023 & 0.0023 & 0.0023
\\
8 &  0.0018&0.0013 & 0.0013 & 0.0013
\\
9 &  0.0014&0.0009 & 0.0009 & 0.0009
\\
10 &  0.0011&0.0006& 0.0006 & 0.0007
\\
11 &  0.0010&0.0004 & 0.0004 & 
\\
\hline
total\tablenotemark[2] & 4.9940 & 4.9864 & 4.9878 & 4.988(2) 
\\
 & & & 4.9877\tablenotemark[3]&
\end{tabular}
\end{ruledtabular}

\tablenotetext[1]{Calculated as ``shell, $N=500$'' plus the model correction $\delta_\text{2pF}$ (See Table \ref{tab:models}).}
\tablenotetext[2]{Sum up to $\abs{\kappa} = 10$ plus $\kappa$-tail extrapolation}
\tablenotetext[3]{Ref. \cite{Yerokhin2015}}

\end{table}

\subsection{Basis extrapolation and error estimation}\label{sec_basis_extr}

\begin{figure}[!htb]
    \centering
    \includegraphics[width=1\linewidth]{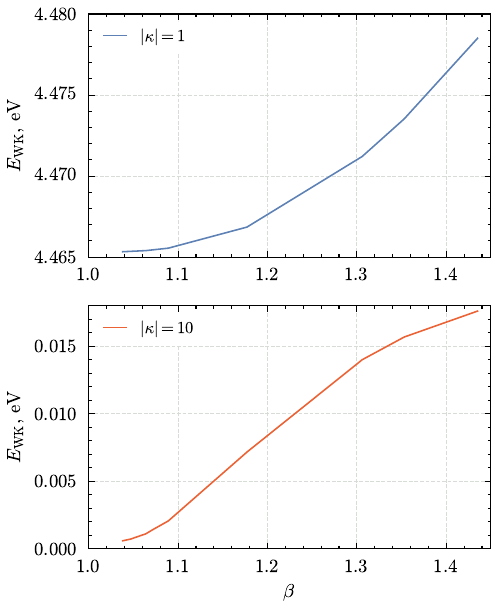}
    \caption{The values of the $E_\text{WK}$ partial components as functions of $\beta$. 
    }
    \label{fig:beta_plot_k1}
\end{figure}

\begin{figure}[!htb]
    \centering
    \includegraphics[width=1\linewidth]{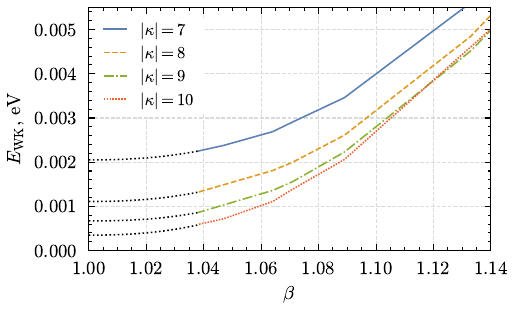}
    \caption{The values of the $E_\text{WK}$ partial components as functions of $\beta$. The black dotted lines show the extrapolation, Eq.~\eqref{power_extr}. 
    }
    \label{fig:beta_plot}
\end{figure}

Since the results for the vacuum polarization correction converge as $N$ increases, we can extrapolate the dependence of the energy shift value on the basis size. These values can be approximated by an inverse polynomial series, such as
\begin{equation}
    \hat{E}(N; \vb{a}) = \sum_{i=0}^{i_\text{max}} \frac{a_i}{N^i} ,
\label{inverse_pow}
\end{equation}
where $\hat{E}$ is the estimated energy correction, $a_i$ are the fit weights. In Table \ref{tab:e_wk1} above, we use this extrapolation to check if the energy correction value converges. However, this dependence does not provide a lower bound for the exact value of the energy shift. Another approach was used in Ref.~\cite{Benazzouk2026}. Instead of using $N$, the $\beta$ parameter was taken as a variable, see Eqs.~\eqref{zetas}, \eqref{complete_basis}. For $\beta \to 1$, this gives a complete basis limit. We shall refer to this as ``basis extrapolation''.
A larger basis set size allows us to approach this limit more closely. Let us explore the corresponding plots. In Figs.~\ref{fig:beta_plot_k1} and \ref{fig:beta_plot} we plot the dependence of the $E_\text{WK}$ components on $\beta$. In both figures, the curves monotonically approach respective limits as we move to the left. In Fig. \ref{fig:beta_plot_k1} we plot the results for $\abs{\kappa} = 1$ and $\abs{\kappa} = 10$. 
Although the first curve is convex throughout the plot, allowing a good fit of the data, the latter exhibits an inflection point: the region with $\beta \lesssim 1.1$ (or $N \gtrsim 200$) must be explored to apply a meaningful fit.

Consider the following approximating functions:
\begin{equation}
    \hat{E}(\beta; \vb{a}) = a_0 + \sum_{i=2}^{\text{degree}} a_i (\beta - 1)^i ,
    \label{cubic_extr}
\end{equation}
\begin{equation}
    \hat{E}(\beta; \vb{a}) = a_0 + a_1 (\beta - 1)^{a_2} .
    \label{power_extr}
\end{equation}
We make the realistic assumption that the $E(\beta)$ curve is monotonic in the convergence region and has a derivative of zero at $\beta = 1$. 
We plot the extrapolating curves in Fig.~\ref{fig:beta_plot}, obtained as a power fit, Eq.~\eqref{power_extr}, of the calculated data. However, note that the fitting estimate is lower than the expected correction values (as can be seen by comparing with independent results or by taking other intervals of $E(\beta)$ curve to extrapolate). For example, for $\abs{\kappa} = 7$ the extrapolation Eq.~\eqref{power_extr} yields 0.00205 eV whereas the rightmost column in Table~\ref{tab:e_wk_2pf} gives 0.0023 eV. The cubic fit, Eq.~\eqref{cubic_extr}, yields similar results (e.g. 0.00203 eV for $\abs{\kappa} = 7$); therefore, we do not show it in the plot.
From the monotonicity of $E(\beta)$ function we can deduce the lower and upper bounds for the exact value of the energy shift value,
\begin{equation}
    E(\beta^*) - E'(\beta^*) (\beta^* - 1) \leq E(1) \leq E(\beta^*) ,
\label{lin_extr}
\end{equation}
where $\beta^*$ is the smallest value of $\beta$ for which $E(\beta)$ was obtained.
This linear extrapolation makes minimal assumptions about the $E(\beta)$ curve and provides a conservative estimate of the energy shift, see Table~\ref{tab:error_estim}. For simplicity, we use the values for $N=300$ and $N=500$ from Table~\ref{tab:e_wk_2pf} to calculate the line slope and obtain the values in third column of Table~\ref{tab:error_estim}. We also study how a smaller basis size affects the error estimation. For instance, the values for $N=220$ and $N=300$ are used in the second column of Table~\ref{tab:error_estim}. A more accurate calculation of the line slope reduces the interval insignificantly. In practice, however, this uncertainty estimate is too large. Using only this estimate would lead to discarding results for $\abs{\kappa} > 6$ in Table~\ref{tab:e_wk_2pf}, while in fact results up to $\abs{\kappa} \leq 10$ are in a good agreement with Ref.~\cite{Mandrykina2026}. Alternatively, we can use the fact that the extrapolation Eq.~\eqref{power_extr} gives an estimate lower than the expected correction value. Using this extrapolation as the lower bound, we obtain another error estimate. As we see in Table~\ref{tab:error_estim}, the uncertainty intervals for this estimation method are much smaller. It is important to note that, since these estimates depend on the extrapolation law, the error interval may vary. Additionally, these estimates do not indicate which value within the interval is more likely to be correct. Finally, the actual form of the $E(\beta)$ curve must be taken into account, as is evident from the lower plot in Fig.~\ref{fig:beta_plot_k1}, where the linear extrapolation should not be used for the right part of the plotted values.

\begin{table*}[!htb]
\footnotesize
\caption{\label{tab:error_estim}%
Estimates of error due to incomplete convergence for WK correction, in eV. The shell nucleus model with $r_n = 5.8571$ fm is considered. The maximum basis size $N_\text{max}$ used for extrapolation is indicated. The midpoint and range of the interval are presented. The ratio of the interval range to its midpoint is shown in parentheses. The sum over first 6 and 10 components and their error estimates are shown in the last two rows.
}
\begin{ruledtabular}
\begin{tabular}{lllll}
$\abs{\kappa}$  &Eq.~\eqref{lin_extr}, $N_\text{max} = 300$ &  Eq.~\eqref{lin_extr}, $N_\text{max} = 500$ &Eq.~\eqref{power_extr}, $N_\text{max} = 300$ & Eq.~\eqref{power_extr}, $N_\text{max} = 500$ \\

\hline
1   &4.46521 $\pm$ 0.00019 (0.00\%)& 4.46528 $\pm$ 0.00005 (0.00\%)     &4.46535 $\pm$ 0.00005 (0.00\%)
& 4.46532 $\pm$ 0.00001 (0.00\%)   \\
2   &0.39333 $\pm$ 0.00037 (0.09\%)& 0.39344 $\pm$ 0.00011 (0.03\%)     &0.39358 $\pm$ 0.00011 (0.03\%)
& 0.39353 $\pm$ 0.00002 (0.01\%)   \\
3   &0.08114 $\pm$ 0.00054 (0.66\%)& 0.08132 $\pm$ 0.00015 (0.19\%)     &0.08152 $\pm$ 0.00016 (0.20\%)
& 0.08143 $\pm$ 0.00003 (0.04\%)   \\
4   &0.02451 $\pm$ 0.00065 (2.66\%)& 0.02468 $\pm$ 0.00020 (0.83\%)     &0.02497 $\pm$ 0.00019 (0.77\%)
& 0.02483 $\pm$ 0.00005 (0.21\%)   \\
5   &0.00918 $\pm$ 0.00077 (8.36\%)& 0.00936 $\pm$ 0.00025 (2.65\%)     &0.00970 $\pm$ 0.00024 (2.51\%)
& 0.00955 $\pm$ 0.00006 (0.61\%)   \\
6   &0.00383 $\pm$ 0.00091 (23.7\%)& 0.00408 $\pm$ 0.00028 (6.80\%)     &0.00441 $\pm$ 0.00032 (7.32\%)
& 0.00431 $\pm$ 0.00006 (1.37\%)   \\
7   &0.00170 $\pm$ 0.00099 (57.8\%)& 0.00193 $\pm$ 0.00032 (16.7\%)&0.00234 $\pm$ 0.00035 (15.1\%)& 0.00215 $\pm$ 0.00010 (4.62\%)   \\
8   &0.00079 $\pm$ 0.00102 (130\%)& 0.00096 $\pm$ 0.00036 (37.2\%)&0.00145 $\pm$ 0.00036 (25.2\%)& 0.00122 $\pm$ 0.00010 (8.63\%)   \\
9   &0.00025 $\pm$ 0.00111 (451\%)& 0.00049 $\pm$ 0.00037 (73.9\%)&0.00092 $\pm$ 0.00044 (47.3\%)& 0.00077 $\pm$ 0.00009 (12.3\%)\\
10  &-0.00011 $\pm$ 0.00122 (1147\%)& 0.00021 $\pm$ 0.00038 (180\%)&0.00056 $\pm$ 0.00055 (99.0\%)& 0.00047 $\pm$ 0.00012 (25.6\%)\\
11  &-0.00020 $\pm$ 0.00123 (618\%)& 0.00001 $\pm$ 0.00043 (4873\%)&0.00047 $\pm$ 0.00056 (120\%)& 0.00032 $\pm$ 0.00012 (39.1\%)\\
\hline
Sum $\abs{\kappa} \leq 6$  &4.97720 $\pm$ 0.00343 (0.07\%)& 4.97815 $\pm$ 0.00104 (0.02\%)  &4.97953 $\pm$ 0.00109 (0.02\%)& 4.97897 $\pm$ 0.00023 (0.00\%)  \\
Sum $\abs{\kappa} \leq 10$  &4.97983 $\pm$ 0.00777 (0.16\%)& 4.98175 $\pm$ 0.00247 (0.05\%)  &4.98480 $\pm$ 0.00279 (0.06\%)& 4.98357 $\pm$ 0.00065 (0.01\%) 
 \end{tabular}
\end{ruledtabular}

\end{table*}

\subsection{Screening effect}

\begin{figure}[!htb]
    \centering
    \includegraphics[width=0.8\linewidth]{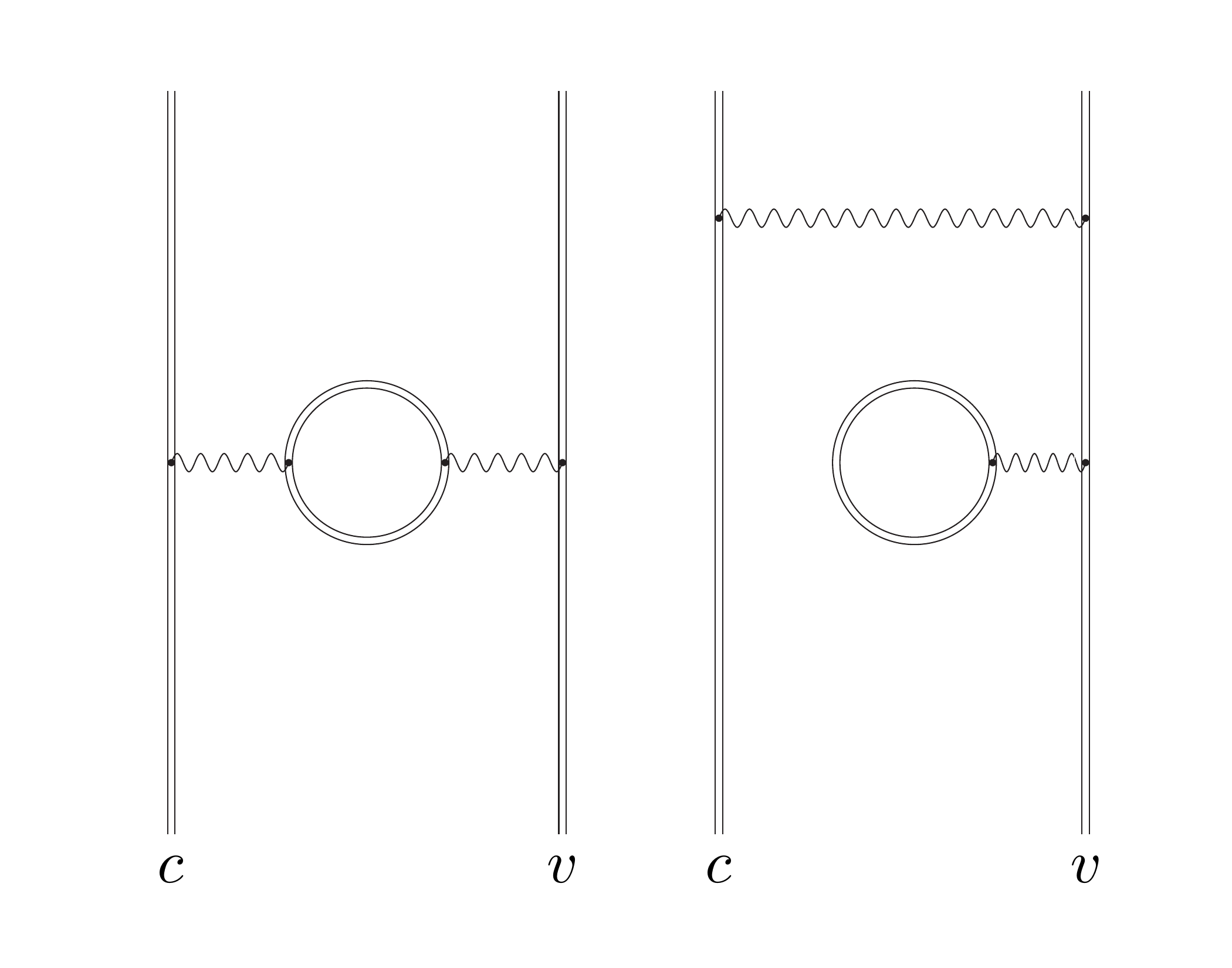}
    \caption{Diagrams contributing to the vacuum-polarization screening effect. 
    }
    \label{fig:screen_diagram}
\end{figure}

Let us examine the application of the Gaussian-basis-set method to calculations involving lithium-like ions. The screening effect can be described by the diagrams in Fig.~\ref{fig:screen_diagram}.
As shown in Ref.~\cite{Persson1993}, we can find the ``electron core'' contribution (direct part of the left diagram in Fig.~\ref{fig:screen_diagram}) by evaluating the vacuum polarization correction in a potential formed by nucleus with an account of the $1s$ electron core.

The screening potential from the $1s^2$ charge density can be considered as a modification of the original potential of the bare nucleus. 
The electron core density and the resulting potential are
\begin{equation}
    \rho_c(r) = \sum_a (2j_a + 1) \left(P_a^2(r) + Q_a^2(r)\right)
\end{equation}
\begin{equation}
    V_\text{scr}(r) = \alpha\int_0^\infty \dd r' \frac{1}{r_>} \rho_c(r')
\label{core_pot}
\end{equation}
where $r_> = \max (r, r')$, symbols $a$ denote the electron core states (e.g. $1s_{1/2}$ in lithium-like ions).
When calculating the $\rho(\delta)$ term in Eq.~\eqref{c_sym} without the electron core, in fact we multiply the $V$ term in the matrix elements of the Hamiltonian by $\delta / Z$. Hence, we can do exactly the same with the modified potential $V_\text{nucl} + V_\text{scr}$. For example, for the $\rho_\kappa(r,\delta)$ term we have the following expression for the potential in Hamiltonian, Eq.~\eqref{dirac_eq_2}:
\begin{equation}
    V = \frac{\delta}{Z} \left( V_\text{nucl} + V_\text{scr} \right) .
\label{scr_add}
\end{equation}
Using this approach, we find $\Delta E_\text{scr} = -0.0044$ eV for the electron core part of the direct screening for the $2s_{1/2}$ state of uranium ion with $r_n = 5.860$ fm and uniform (sphere) charge distribution. We used basis sets with $N = 100$ and $N = 150$. This result can be compared with $\Delta E_\text{scr} = -0.0043$ eV of Persson et al. \cite{Persson1993}. To calculate the contribution from the direct part of the right diagram in Fig.~\ref{fig:screen_diagram} we take the expectation value with the screened states, but use the WK potential generated only from the nuclear potential. In this way, we obtain $\Delta E_\text{scr} = -0.0322$ eV, to be compared with $\Delta E_\text{scr} = -0.032$ eV from Ref.~\cite{Persson1993}. A careful treatment of the remaining contributions (e.g. exchange Coulomb and Breit interaction) leads to the complete value of the Wichmann-Kroll correction in lithium-like ions, but this remaining value is usually small and can be neglected \cite{Sapirstein2003}.

Alternatively, the value of the screening correction can be obtained using the screening potentials (sometimes called non-Coulomb). Although for the Uehling correction this approach is not sufficient and the screening effect should be considered in more detail \cite{Sapirstein2003}, for the Wichmann-Kroll correction it can be calculated with good accuracy. The core-Hartree screening potential is obtained by iteratively finding electron wave functions in the field of the screening potential formed by the electron core, see Eq.~\eqref{core_pot}. The resulting potential is then used to find the vacuum polarization correction.
In Ref.~\cite{Sapirstein2003} the Kohn-Sham potential was used to find the vacuum polarization correction in lithium-like ions. This potential is given 
\begin{equation}
    V_\text{scr}(r) = \alpha\int_0^\infty \dd r' \frac{1}{r_>} \rho_t(r') - \frac{\alpha}{r}\frac{2}{3} \left(\frac{81}{32\pi^2}r \rho_t(r)\right)^{1/3},
\end{equation}
\begin{equation}
    \rho_t(r) = P_v^2(r) + Q_v^2(r) + \rho_c(r),
\end{equation}
where $v$ denotes the state of the valence electron.

In order to use these potentials in our computational scheme, one only needs to add them to the nuclear potential and take expectation values with screened states. The results are shown in Table \ref{tab:screen} for uranium with $r_n = 5.860$ fm and the Fermi model distribution. We use the values of the first three partial components to calculate the screening effect here. 
\begin{table}[!htb]
\footnotesize
\caption{\label{tab:screen}%
Screening effect in lithium-like uranium for $2s_{1/2}$ state. The effect is calculated using core-Hartree (CH) and Kohn-Sham (KS) potentials.
}
\begin{ruledtabular}
\begin{tabular}{c|cc}
$N$ & CH & KS \\
\hline
$50$  & -0.0367 & -0.0313 \\
$100$ & -0.0362 & -0.0309 \\
$150$ & -0.0361 & -0.0308 \\
$200$ & -0.0361 & -0.0308 \\
\hline
\cite{Sapirstein2003} & -0.036 & -0.031 
\\\end{tabular}
\end{ruledtabular}
\end{table}

This effect can also be calculated using the Dirac-Hartree-Fock method. In a recent work, Ref.~\cite{Hayat2026arxiv}, in this approach the Gaussian basis was used to calculate the screening correction for the vacuum polarization effect. The authors found that for uranium $\Delta E_\text{scr} = -0.0389$ eV.

\section{Conclusion}\label{section_v}

In this work, we calculated the Wichmann-Kroll vacuum polarization correction using the Gaussian basis set with a number of basis functions up to 500 basis functions. We have found that the results for the main and several of the first partial components of this correction converge very closely to the numbers offered by Green's function integration. However, convergence becomes slower for larger values of $\abs{\kappa}$. Nevertheless, this level of accuracy is sufficient to obtain the results for $\abs{\kappa} \leq 10$ that are very close to those of the Green's function method, as shown in Table \ref{tab:e_wk_2pf}. Adding the $\kappa$-tail nearly reproduces the result of Ref.~\cite{Yerokhin2015}.

We have discussed basis extrapolation techniques and their limitations. 
Our large basis calculations allowed us to carefully investigate the dependence of the energy correction due to the basis size. In terms of parameter $\beta$ we reached the value of $\beta = 1.038$, which is close to the complete basis limit.
Our analysis indicates that a simple fitting law can only be used for rough estimates, especially for higher values of $\abs{\kappa}$.
Rather than using extrapolation to predict the exact values, we propose to use extrapolation to estimate errors due to incomplete convergence in the obtained results independently of the other computational methods. 

We studied the effect of the various nuclear models and showed that these values converge more quickly, compared to the convergence of the absolute value of the energy shift. This enables us to calculate the model corrections using a smaller size of the basis set.

Finally, we have discussed calculating the vacuum-polarization screening effect in lithium-like ions and have presented the results for the Wichmann-Kroll correction for the $2s_{1/2}$ electron. The results are shown to be very close to those of the previous works, and their rapid convergence has been demonstrated.

\begin{acknowledgments}
We are grateful to J. Sommerfeldt and V. Agababaev for helpful discussions.

This work is supported by the Ministry of Science and Higher Education of the Russian Federation (Project FSER-2025-0012)
and the Foundation for the Advancement of Theoretical Physics and Mathematics “BASIS” (Grant No. 25-1-5-126-1).

\end{acknowledgments}

\bigbreak

\bibliography{main.bib}

@article{Benazzouk2026,
  title = {Wichmann-Kroll vacuum polarization density in a finite Gaussian basis set},
  author = {Benazzouk, Ryan and Salman, Maen and Saue, Trond},
  journal = {Phys. Rev. A},
  volume = {113},
  issue = {3},
  pages = {032813},
  numpages = {19},
  year = {2026},
  month = {Mar},
  publisher = {American Physical Society},
  doi = {10.1103/wbz7-czxx},
  url = {https://link.aps.org/doi/10.1103/wbz7-czxx}
}

@article{Sommerfeldt2026,
  title = {All-order Wichmann-Kroll contribution in heavy electronic and exotic atoms},
  author = {Sommerfeldt, Jonas and Indelicato, Paul},
  journal = {Phys. Rev. A},
  volume = {113},
  issue = {2},
  pages = {022806},
  numpages = {8},
  year = {2026},
  month = {Feb},
  publisher = {American Physical Society},
  doi = {10.1103/ylft-n4f9},
  url = {https://link.aps.org/doi/10.1103/ylft-n4f9}
}

@misc{Mandrykina2026,
      title={Wichmann-Kroll Correction in Muonic Atoms and Hydrogen-Like Electronic Ions: a Comparative Study of Two Methods}, 
      author={Zoia A. Mandrykina and Zewen Sun and Natalia S. Oreshkina},
      year={2026},
      eprint={2601.22979},
      archivePrefix={arXiv},
      primaryClass={physics.atom-ph},
      url={https://arxiv.org/abs/2601.22979}, 
}

@article{Parpia1992,
  title = {Relativistic basis-set calculations for atoms with Fermi nuclei},
  author = {Parpia, F. A. and Mohanty, A. K.},
  journal = {Phys. Rev. A},
  volume = {46},
  issue = {7},
  pages = {3735--3745},
  numpages = {0},
  year = {1992},
  month = {Oct},
  publisher = {American Physical Society},
  doi = {10.1103/PhysRevA.46.3735},
  url = {https://link.aps.org/doi/10.1103/PhysRevA.46.3735}
}

@article{Ivanov2024,
  title = {Vacuum-polarization Wichmann-Kroll correction in the finite-basis-set approach},
  author = {Ivanov, V. K. and Baturin, S. S. and Glazov, D. A. and Volotka, A. V.},
  journal = {Phys. Rev. A},
  volume = {110},
  issue = {3},
  pages = {032815},
  numpages = {11},
  year = {2024},
  month = {Sep},
  publisher = {American Physical Society},
  doi = {10.1103/PhysRevA.110.032815},
  url = {https://link.aps.org/doi/10.1103/PhysRevA.110.032815}
}

@article{Ivanov2025,
  title = {Sturmian basis set for the Dirac equation with finite nuclear size: Application to polarizability, Zeeman and hyperfine splitting, and vacuum polarization},
  author = {Ivanov, V. K. and Glazov, D. A. and Volotka, A. V.},
  journal = {Phys. Rev. A},
  volume = {112},
  issue = {2},
  pages = {022816},
  numpages = {11},
  year = {2025},
  month = {Aug},
  publisher = {American Physical Society},
  doi = {10.1103/tf1t-wj6w},
  url = {https://link.aps.org/doi/10.1103/tf1t-wj6w}
}

@article{Yerokhin2015,
    author = {Yerokhin, V. A. and Shabaev, V. M.},
    title = "{Lamb Shift of n = 1 and n = 2 States of Hydrogen-like Atoms, 1 ≤ Z ≤ 110}",
    journal = {Journal of Physical and Chemical Reference Data},
    volume = {44},
    number = {3},
    pages = {033103},
    year = {2015},
    month = {08},
    issn = {0047-2689},
    doi = {10.1063/1.4927487},
    url = {https://doi.org/10.1063/1.4927487},
}

@article{Indelicato2019,
doi = {10.1088/1361-6455/ab42c9},
url = {https://doi.org/10.1088/1361-6455/ab42c9},
year = {2019},
month = {nov},
publisher = {IOP Publishing},
volume = {52},
number = {23},
pages = {232001},
author = {Indelicato, P},
title = {QED tests with highly charged ions},
journal = {Journal of Physics B: Atomic, Molecular and Optical Physics}
}

@article{Beier2000,
title = {The $g_j$ factor of a bound electron and the hyperfine structure splitting in hydrogenlike ions},
journal = {Physics Reports},
volume = {339},
number = {2},
pages = {79-213},
year = {2000},
issn = {0370-1573},
doi = {https://doi.org/10.1016/S0370-1573(00)00071-5},
url = {https://www.sciencedirect.com/science/article/pii/S0370157300000715},
author = {Thomas Beier}
}

@article{Manakov1989,
  title={Vacuum polarization by a strong coulomb field and its contribution to the spectra of multiply-charged ions},
  author={Manakov, NL and Nekipelov, AA and Fainshtein, AG},
  journal={Zh. Eksp. Teor. Fiz},
  volume={95},
  pages={1167--1177},
  year={1989}
}

@article{Mohr1974,
title = {Numerical evaluation of the 1S12-state radiative level shift},
journal = {Annals of Physics},
volume = {88},
number = {1},
pages = {52-87},
year = {1974},
issn = {0003-4916},
doi = {https://doi.org/10.1016/0003-4916(74)90399-6},
url = {https://www.sciencedirect.com/science/article/pii/0003491674903996},
author = {Peter J Mohr}
}

@misc{Hayat2026arxiv,
      title={Wichmann-Kroll vacuum polarization correction to lithium-like systems in a Gaussian basis set}, 
      author={Haisum Hayat and Harry M. Quiney},
      year={2026},
      eprint={2604.23660},
      archivePrefix={arXiv},
      primaryClass={physics.atom-ph},
      url={https://arxiv.org/abs/2604.23660}, 
}

@article{Ferenc2025,
  title = {Gaussian-basis-set approach to one-loop self-energy},
  author = {Ferenc, D\'avid and Salman, Maen and Saue, Trond},
  journal = {Phys. Rev. A},
  volume = {111},
  issue = {4},
  pages = {L040802},
  numpages = {6},
  year = {2025},
  month = {Apr},
  publisher = {American Physical Society},
  doi = {10.1103/PhysRevA.111.L040802},
  url = {https://link.aps.org/doi/10.1103/PhysRevA.111.L040802}
}

@article{Sapirstein2003,
  title = {Vacuum polarization calculations for hydrogenlike and alkali-metal-like ions},
  author = {Sapirstein, J. and Cheng, K. T.},
  journal = {Phys. Rev. A},
  volume = {68},
  issue = {4},
  pages = {042111},
  numpages = {10},
  year = {2003},
  month = {Oct},
  publisher = {American Physical Society},
  doi = {10.1103/PhysRevA.68.042111},
  url = {https://link.aps.org/doi/10.1103/PhysRevA.68.042111}
}

@article{Klahn1985,
    author = {Klahn, Bruno},
    title = {A generalization of the Müntz–Szász theorem to floating exponents with applications to Gauss‐ and Slater‐type functions},
    journal = {The Journal of Chemical Physics},
    volume = {83},
    number = {11},
    pages = {5749-5753},
    year = {1985},
    month = {12},
    issn = {0021-9606},
    doi = {10.1063/1.449651},
    url = {https://doi.org/10.1063/1.449651}
}

@article{mohr2024codata,
  title = {CODATA recommended values of the fundamental physical constants: 2022},
  author = {Mohr, Peter J. and Newell, David B. and Taylor, Barry N. and Tiesinga, Eite},
  journal = {Rev. Mod. Phys.},
  volume = {97},
  issue = {2},
  pages = {025002},
  numpages = {62},
  year = {2025},
  month = {Apr},
  publisher = {American Physical Society},
  doi = {10.1103/RevModPhys.97.025002},
  url = {https://link.aps.org/doi/10.1103/RevModPhys.97.025002}
}

@article{Moskovkin2004,
  title = {$g$ factor of hydrogenlike ions with nonzero nuclear spin},
  author = {Moskovkin, D. L. and Oreshkina, N. S. and Shabaev, V. M. and Beier, T. and Plunien, G. and Quint, W. and Soff, G.},
  journal = {Phys. Rev. A},
  volume = {70},
  issue = {3},
  pages = {032105},
  numpages = {11},
  year = {2004},
  month = {Sep},
  publisher = {American Physical Society},
  doi = {10.1103/PhysRevA.70.032105},
  url = {https://link.aps.org/doi/10.1103/PhysRevA.70.032105}
}

@article{Salman2023,
  title = {Calculating the many-potential vacuum polarization density of the Dirac equation in the finite-basis approximation},
  author = {Salman, Maen and Saue, Trond},
  journal = {Phys. Rev. A},
  volume = {108},
  issue = {1},
  pages = {012808},
  numpages = {13},
  year = {2023},
  month = {Jul},
  publisher = {American Physical Society},
  doi = {10.1103/PhysRevA.108.012808},
  url = {https://link.aps.org/doi/10.1103/PhysRevA.108.012808}
}

@article{Shabaev_DKB,
  title = {Dual Kinetic Balance Approach to Basis-Set Expansions for the Dirac Equation},
  author = {Shabaev, V. M. and Tupitsyn, I. I. and Yerokhin, V. A. and Plunien, G. and Soff, G.},
  journal = {Phys. Rev. Lett.},
  volume = {93},
  issue = {13},
  pages = {130405},
  numpages = {4},
  year = {2004},
  month = {Sep},
  publisher = {American Physical Society},
  doi = {10.1103/PhysRevLett.93.130405},
  url = {https://link.aps.org/doi/10.1103/PhysRevLett.93.130405}
}

@article{Mohr1998,
title = {QED corrections in heavy atoms},
journal = {Physics Reports},
volume = {293},
number = {5},
pages = {227-369},
year = {1998},
issn = {0370-1573},
doi = {https://doi.org/10.1016/S0370-1573(97)00046-X},
url = {https://www.sciencedirect.com/science/article/pii/S037015739700046X},
author = {Peter J. Mohr and Günter Plunien and Gerhard Soff}
}

@article{Johnson1988,
  title = {Finite basis sets for the Dirac equation constructed from B splines},
  author = {Johnson, W. R. and Blundell, S. A. and Sapirstein, J.},
  journal = {Phys. Rev. A},
  volume = {37},
  issue = {2},
  pages = {307--315},
  numpages = {0},
  year = {1988},
  month = {Jan},
  publisher = {American Physical Society},
  doi = {10.1103/PhysRevA.37.307},
  url = {https://link.aps.org/doi/10.1103/PhysRevA.37.307}
}

@article{Sapirstein1996,
    doi = {10.1088/0953-4075/29/22/005},
    url = {https://dx.doi.org/10.1088/0953-4075/29/22/005},
    year = {1996},
    month = {nov},
    publisher = {},
    volume = {29},
    number = {22},
    pages = {5213},
    author = {J Sapirstein and  W R Johnson},
    title = {The use of basis splines in theoretical atomic physics},
    journal = {Journal of Physics B: Atomic, Molecular and Optical Physics}
}

@article{Rinker1975,
  title = {Vacuum polarization in strong, realistic electric fields},
  author = {Rinker, G. A. and Wilets, L.},
  journal = {Phys. Rev. A},
  volume = {12},
  issue = {3},
  pages = {748--762},
  numpages = {0},
  year = {1975},
  month = {Sep},
  publisher = {American Physical Society},
  doi = {10.1103/PhysRevA.12.748},
  url = {https://link.aps.org/doi/10.1103/PhysRevA.12.748}
}

@article{Wichmann1956,
  title = {Vacuum Polarization in a Strong Coulomb Field},
  author = {Wichmann, Eyvind H. and Kroll, Norman M.},
  journal = {Phys. Rev.},
  volume = {101},
  issue = {2},
  pages = {843--859},
  numpages = {0},
  year = {1956},
  month = {Jan},
  publisher = {American Physical Society},
  doi = {10.1103/PhysRev.101.843},
  url = {https://link.aps.org/doi/10.1103/PhysRev.101.843}
}

@article{Persson1993,
  title = {Accurate vacuum-polarization calculations},
  author = {Persson, Hans and Lindgren, Ingvar and Salomonson, Sten and Sunnergren, Per},
  journal = {Phys. Rev. A},
  volume = {48},
  issue = {4},
  pages = {2772--2778},
  numpages = {0},
  year = {1993},
  month = {Oct},
  publisher = {American Physical Society},
  doi = {10.1103/PhysRevA.48.2772},
  url = {https://link.aps.org/doi/10.1103/PhysRevA.48.2772}
}

@article{Soff1988,
  title = {Vacuum polarization in a strong external field},
  author = {Soff, Gerhard and Mohr, Peter J.},
  journal = {Phys. Rev. A},
  volume = {38},
  issue = {10},
  pages = {5066--5075},
  numpages = {0},
  year = {1988},
  month = {Nov},
  publisher = {American Physical Society},
  doi = {10.1103/PhysRevA.38.5066},
  url = {https://link.aps.org/doi/10.1103/PhysRevA.38.5066}
}

@article{Grant2000,
  title = {Rayleigh-Ritz approximation of the Dirac operator in atomic and molecular physics},
  author = {Grant, I. P. and Quiney, H. M.},
  journal = {Phys. Rev. A},
  volume = {62},
  issue = {2},
  pages = {022508},
  numpages = {14},
  year = {2000},
  month = {Jul},
  publisher = {American Physical Society},
  doi = {10.1103/PhysRevA.62.022508},
  url = {https://link.aps.org/doi/10.1103/PhysRevA.62.022508}
}

@article{Sturm2014,
    author = {Sturm, Sven and Köhler, F and Zatorski, Jacek and Wagner, Annekatrin and Harman, Zoltán and Werth, G. and Quint, Werner and Keitel, C and Blaum, K},
    year = {2014},
    month = {02},
    pages = {},
    title = {High-precision measurement of the atomic mass of the electron},
    volume = {506},
    journal = {Nature},
    doi = {10.1038/nature13026}
}

@article{Shabaev2006,
  title = {$g$-Factor of Heavy Ions: A New Access to the Fine Structure Constant},
  author = {Shabaev, V. M. and Glazov, D. A. and Oreshkina, N. S. and Volotka, A. V. and Plunien, G. and Kluge, H.-J. and Quint, W.},
  journal = {Phys. Rev. Lett.},
  volume = {96},
  issue = {25},
  pages = {253002},
  numpages = {4},
  year = {2006},
  month = {Jun},
  publisher = {American Physical Society},
  doi = {10.1103/PhysRevLett.96.253002},
  url = {https://link.aps.org/doi/10.1103/PhysRevLett.96.253002}
}

@article{Yerokhin2016,
  title = {$g$ Factor of Light Ions for an Improved Determination of the Fine-Structure Constant},
  author = {Yerokhin, V. A. and Berseneva, E. and Harman, Z. and Tupitsyn, I. I. and Keitel, C. H.},
  journal = {Phys. Rev. Lett.},
  volume = {116},
  issue = {10},
  pages = {100801},
  numpages = {5},
  year = {2016},
  month = {Mar},
  publisher = {American Physical Society},
  doi = {10.1103/PhysRevLett.116.100801},
  url = {https://link.aps.org/doi/10.1103/PhysRevLett.116.100801}
}

@article{Gyulassy1974,
  title = {Vacuum Polarization in Heavy-Ion Collisions},
  author = {Gyulassy, Miklos},
  journal = {Phys. Rev. Lett.},
  volume = {33},
  issue = {15},
  pages = {921--925},
  numpages = {0},
  year = {1974},
  month = {Oct},
  publisher = {American Physical Society},
  doi = {10.1103/PhysRevLett.33.921},
  url = {https://link.aps.org/doi/10.1103/PhysRevLett.33.921}
}

@article{Yerokhin2019,
    author = {Yerokhin, Vladimir A. and Pachucki, Krzysztof and Patkóš, Vojtěch},
    title = {Theory of the Lamb Shift in Hydrogen and Light Hydrogen-Like Ions},
    journal = {Annalen der Physik},
    volume = {531},
    number = {5},
    pages = {1800324},
    doi = {https://doi.org/10.1002/andp.201800324},
    url = {https://onlinelibrary.wiley.com/doi/abs/10.1002/andp.201800324},
    year = {2019}
}

@article{Volotka2013,
    author = {Volotka, Andrey V. and Glazov, Dmitry A. and Plunien, Günter and Shabaev, Vladimir M.},
    title = {Progress in quantum electrodynamics theory of highly charged ions},
    journal = {Annalen der Physik},
    volume = {525},
    number = {8-9},
    pages = {636-646},
    doi = {https://doi.org/10.1002/andp.201300079},
    url = {https://onlinelibrary.wiley.com/doi/abs/10.1002/andp.201300079},
    year = {2013}
}

@article{Gyulassy1975,
    title = {Higher order vacuum polarization for finite radius nuclei},
    journal = {Nuclear Physics A},
    volume = {244},
    number = {3},
    pages = {497-525},
    year = {1975},
    issn = {0375-9474},
    doi = {https://doi.org/10.1016/0375-9474(75)90554-0},
    url = {https://www.sciencedirect.com/science/article/pii/0375947475905540},
    author = {Miklos Gyulassy}
}

@article{Rinker1973,
  title = {Vacuum Polarization in High-$Z$, Finite-Size Nuclei},
  author = {Rinker, G. A. and Wilets, L.},
  journal = {Phys. Rev. Lett.},
  volume = {31},
  issue = {26},
  pages = {1559--1562},
  numpages = {0},
  year = {1973},
  month = {Dec},
  publisher = {American Physical Society},
  doi = {10.1103/PhysRevLett.31.1559},
  url = {https://link.aps.org/doi/10.1103/PhysRevLett.31.1559}
}

\end{document}